\documentclass[3p]{elsarticle}

\usepackage{lineno,hyperref}
\usepackage{graphicx}
\usepackage{amsmath}
\usepackage{natbib}
\usepackage{color}
\usepackage[export]{adjustbox}

\makeatletter
\def\ps@pprintTitle{%
   \let\@oddhead\@empty
   \let\@evenhead\@empty
   \def\@oddfoot{\reset@font\hfil\thepage\hfil}
   \let\@evenfoot\@oddfoot
}
\makeatother

\begin{document}

\begin{frontmatter}

\title{Sensitivity to neutrino-antineutrino transitions for boron neutrinos}

\author[SYSU]{S.J. Li}
\author[SYSU,Tsinghua]{J.J. Ling\corref{mycorrespondingauthor}}
\ead{lingjj5@mail.sysu.edu.cn}
\author[SYSU]{N. Raper}
\author[SYSU]{M.V. Smirnov\corref{mycorrespondingauthor}}
\ead{gear8mike@gmail.com}

\address[SYSU]{Department of Physics, Sun Yat-Sen University,\\Guangzhou 510275, China}
\address[Tsinghua]{Key Laboratory of Particle \& Radiation Imaging (Tsinghua University),\\
Ministry of Education, Beijing 10084, China}

\cortext[mycorrespondingauthor]{Corresponding authors}

\begin{abstract}
	Neutrino-antineutrino conversion is an important new physics process. The observation of this phenomenon could indicate total lepton number violation and potential CPT-violation.
	Searching for the appearance of electron antineutrinos from solar neutrinos from $\rm^8B$ decay allows us to hunt for this rare process, although it can also be explained by other mechanisms or hypotheses.
	This analysis examines the capabilities of observing neutrino-antineutrino transition from $\rm^8B$ unoscillated solar neutrinos using different liquid scintillator detector configurations.
	High energy reactor neutrinos and atmospheric neutrinos are the two dominant background sources.
	Large volume liquid scintillator detectors with deep underground shielding, placed far away from reactors and with capabilities of pulse shaped discrimination will significantly increase the search sensitivity.
	It is demonstrated that for the next generation of large liquid scintillator detectors being planned or under construction, the sensitivity to the average probability of neutrino-antineutrino transitions can reach $10^{-6}$, which is an order of magnitude better than the current best experimental limits.
\end{abstract}

\begin{keyword}
Solar neutrinos\sep neutrino-antineutrino transitions\sep  liquid scintillator
\end{keyword}

\end{frontmatter}


\section{Introduction}
\label{sec:intro}

The conservation of lepton numbers, which is not connected with gauge symmetry, has an accidental character in the Standard Model (SM).
In principle physics beyond the SM does not obligate conversation of lepton numbers.
Additionally, there is a strong bond between lepton number violation in the presence of CP-violation and leptogenesis as a possible explanation for baryon asymmetry.
After the successful discovery of neutrino oscillations, which has demonstrated that neutrinos are massive particles and the violation of flavor lepton numbers ($L_e, L_\mu, L_\tau$)~\cite{osc_obs},
it became clear that the SM has to be extended.
However, flavor lepton number is conserved in processes with charged leptons. For example, neutrinoless muon decay $\mu^\pm\rightarrow e^\pm+\gamma$ with branching ratio $>4.2\cdot10^{-13}$ is excluded at 90\% C.L.~\cite{MEG}, which is kinematically allowed.
Furthermore, the violation of total lepton number has not yet been experimentally observed.
Another important issue for the nature of neutrino mass, Dirac or Majorana, has not been resolved yet.
Dirac particles mean that neutrinos and antineutrinos are two different particles. However, if neutrinos are Majorana particles, they are identical to antineutrinos, which means total lepton number violation with $\Delta L=2$.
The current best probe for the Majorana nature of neutrinos is searching for neutrinoless double beta-decay~\cite{double_beta_status}.

To search neutrino-antineutrino transition from solar neutrinos~\cite{Bahcall:1978}-\cite{Langacker} is another possible way of observing total lepton number violation.
The idea was first mentioned by Bruno Pontecorvo in 1957~\cite{Pontecorvo}.
This process involves breaking CPT symmetry which can be interpreted in the framework of the Standard Model Extension (SME)~\cite{Diaz:2009}.
The analysis of CPT-violation for solar neutrinos is discussed in~\cite{Diaz:2016fqd}.
To be noted, solar $\nu \rightarrow \bar\nu$ can also be explained using several exotic ideas.
One is a quite popular theory from the time before neutrino oscillation was observed, spin-flavor precession (SFP).
In a strong magnetic field neutrinos can be converted into antineutrinos if the neutrino magnetic moment is non-zero~\cite{okun}.
This process goes in two stages.
The first stage is when $\nu_e$ turns into $\bar\nu_\mu$ inside the Sun, and the second stage -- $\bar\nu_\mu$ oscillates to $\bar\nu_e$ during propagation to the Earth~\cite{Akhmedov_2}.
Several other options for transitions are the decay of the heaviest neutrino mass eigenstate into the light antineutrino mass eigenstate~\cite{Beacom:2002}, or the oscillation of neutrinos into sterile states and then into antineutrinos~\cite{Smirnov} and etc.
It is worth mentioning that observation of  $\nu\rightarrow\bar\nu$ will not automatically imply the Majorana nature of neutrinos.

To date the experimental upper limits on transition between neutrinos and antineutrinos are set from two solar neutrino measurements: Borexino experiment with $1.3\times10^{-4}$~\cite{borexino_limit} and KamLAND experiment with $5.3\times10^{-5}$ at 90\% C.L.~\cite{kamland_limit}.
Since liquid scintillator (LSc) detectors have proven successful in solar neutrino measurements which have placed limits on neutrino-antineutrino transitions, this research aims to explore the potential of future LSc detectors to further investigate this phenomenon.

\section{$^8$B neutrinos from the Sun}

\begin{figure}[ht]
\centering \includegraphics[scale=0.4]{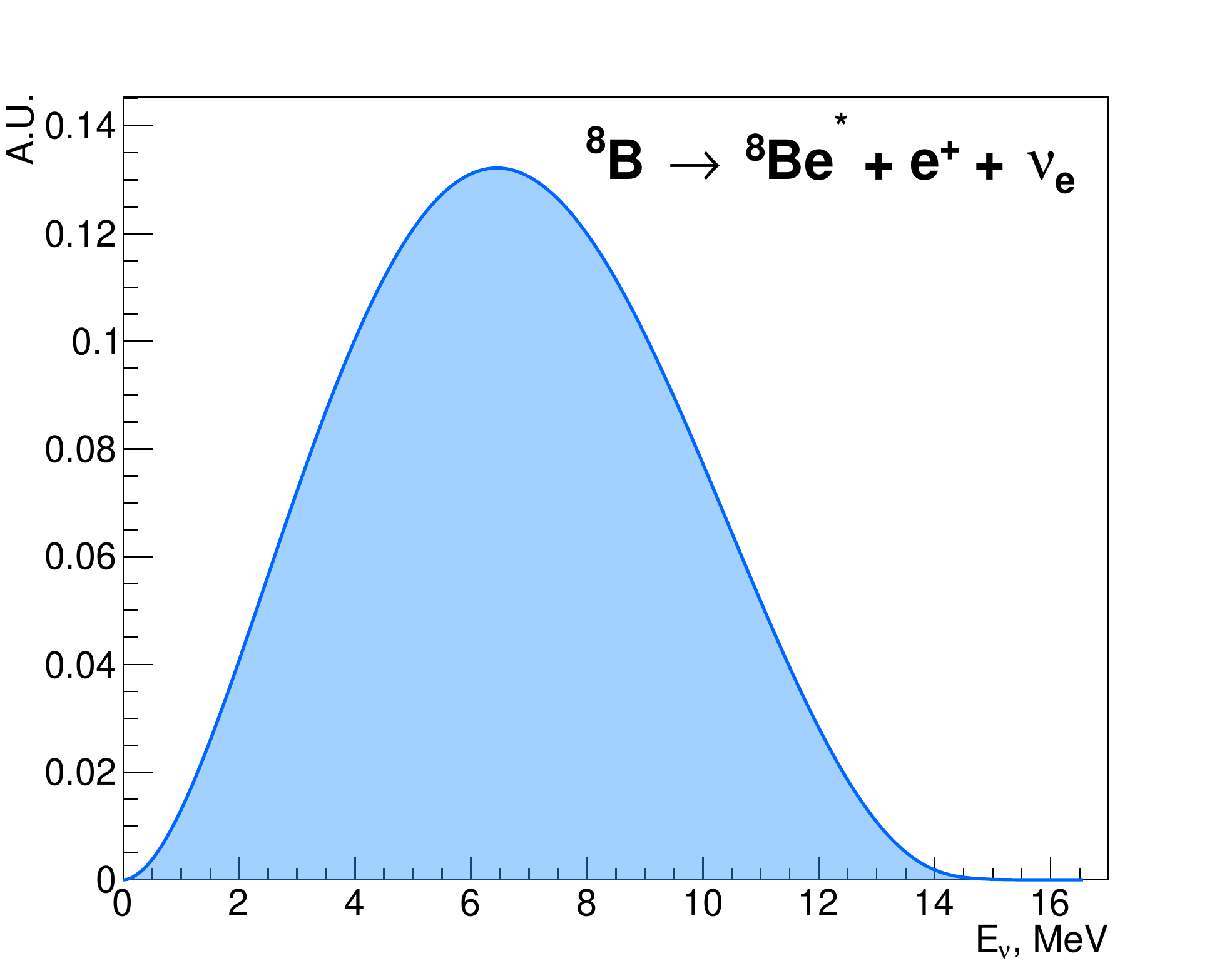}
\caption{\label{fig:1}{The normalized energy spectrum of $^8$B neutrinos with an end point near 15.8 MeV~\cite{8B_spec}.}}
\end{figure}

It is a well-known fact that the Sun emits huge amount of electron neutrinos with energy below 20 MeV. The majority of them are below 2 MeV.
Only boron and hep neutrinos have an energy higher than the threshold of the inverse beta-decay reaction (IBD).
The flux of hep neutrinos is a few orders of magnitude less than the flux of boron neutrinos.
$^8$B is created inside the Sun as part of the {\it pp}-III branch of {\it pp}-chain via the reaction of $^7$Be and a single proton.
Following that $^8$B undergoes beta-plus decay which emits a single electron neutrino.
The energy spectrum of boron neutrinos is continuous with an end point near 15.8 MeV.
The normalized shape of the energy spectrum is shown in Figure\ref{fig:1}.
With the assumption of high metallicity of the Standard Solar Model (SSM), the flux of boron neutrinos is $\rm 5.46 \times 10^6 ~cm^{-2}\cdot s^{-1}$ with a relative uncertainty of 12\%~\cite{solar_model}.

Thus if some parts of the boron solar neutrinos convert into antineutrinos, they can be detected by LSc detectors utilizing the IBD channel.
Boron neutrinos are probably the most convenient source for the observation of neutrino-antineutrino transition.
They have very long propagation time and well known spectra compared with other available neutrino sources. Additionally the flux and energy are convenient for LSc detectors. 
The averaged transition probability between $\nu_e$ and $\bar\nu_e$ for Boron neutrinos can be expressed as:
\begin{equation}
\label{eq_1}
\langle P(\nu_e\to\bar\nu_e)\rangle=\frac{N_{\rm \scriptscriptstyle IBD}}{\Phi\cdot T\cdot n_p\cdot\varepsilon\int\sigma_{\rm \scriptscriptstyle IBD}(E_\nu)S(E_\nu)dE_\nu},
\end{equation}
where $N_{\rm \scriptscriptstyle IBD}$ -- number of detected IBD-events; $\Phi$ -- boron neutrino flux from the Sun; $T$ -- exposure time; $n_p$ -- total number of free protons in the volume of the detector (for LAB LSc $n_p = 6.6 \times 10^{28}~{\rm m^{-3}}$~\cite{lena}); $\varepsilon$ --  efficiency of IBD event selection; $\sigma_{\rm \scriptscriptstyle IBD}(E_\nu)$ -- IBD cross-section~\cite{IBD-xsec}; $S(E_\nu)$ -- shape of the boron neutrino spectrum.

\section{Liquid scintillator types and experimental configurations}
Large active volume, high energy resolution and low energy threshold all make LSc as a popular target material for detecting $\bar{\nu}_e$ through the IBD channel ($\bar{\nu}_e+p \rightarrow e^++n$).
The positron quickly deposits energy and turns into two annihilation gammas, which give the prompt signal.
After $\sim200~{\rm\mu s}$ the neutron is captured by a proton and emits a 2.2 MeV gamma. This unique double-coincidence feature of the IBD signals can largely reduce the detector backgrounds with quite high signal efficiency.

Several different types of large liquid scintillator detectors have been built or proposed for various physics purposes. The KamLAND experiment is located at Kamioka observatory near Toyama, Japan. It has the world largest LSc detector with the target mass of 1 kton. It has an averaged flux-weighted distance of 180 km from the reactors~\cite{kamland_limit}. Borexino is located at Gran Sasso National Laboratory, Italy. It has the world's most radio-pure liquid scintillator with 273 tons target mass~\cite{borexino_limit}. Those two experiments have placed the current best limits on solar neutrino-antineutrino conversion. There are two proposed large scale LSc detectors in China, JUNO~\cite{juno} and Jinping~\cite{Jinping}. JUNO stands for the {\bf J}iangmen {\bf U}nderground {\bf N}eutrino {\bf O}bservatory, which is located in Guangdong Province, south China 53 km away from two powerful reactors. The design calls for a 20 kton liquid scintillator detector with unprecedentedly high energy resolution of $3\%/\sqrt{E({\rm MeV})}$. Upon construction in 2020 it will be the largest LSc detector in the world. Jinping neutrino detector will be located in Jinping Mountain, Sichuan Province, China with maximum overburden around 2400 meters.  It has the lowest muon flux $\simeq2\cdot10^{-10}~\rm{cm^{-2}\cdot s^{-1}}$ compared with other experimental facilities. The Jinping collaboration plans to build a 2 kton detector using slow scintillator. This delays the scintillation process and thus separates from the Cherenkov light. This can significantly increase the background rejection capability using the particle identification method (PID)~\cite{slow}.

Given those different LSc types and detector configurations, this analysis considers two different experimental setups: reactor-based experiment (RE) which usually has relatively high IBD rates from reactors and other background sources, and low background experiment (LBE) which is located at deep underground and far away from reactors with quite low background.
We also consider two options for the LBE target scintillator: normal and slow LSc. It should be mentioned that slow LSc usually does not work with a RE, because the event rate would be too high.

\section{IBD-background and the choice of energy windows}
Since our analysis is based on a simple counting model, the most significant issues are the estimation of backgrounds and the choice of energy windows for RE and LBE.
Requiring coincidence between prompt and delayed signals for IBD selection allows single signals from background to be suppressed dramatically.
However this does not reject background completely.
The main sources of background are listed below:
\begin{itemize}
\item \textbf{\emph{Reactor neutrino and geo-neutrino}}: Reactor neutrino and geo-neutrino are generated from the fission of Uranium and Putonium isotopes in the reactors, and beta decays of Uranium and Thorium inside the terrestrial crust, respectively. Those are the intrinsic electron anti-neutrino backgrounds.
	Reactor neutrino energy can reach $\sim$10 MeV~\cite{reactorbkg} and most of the geo-neutrinos are below 3.4 MeV~\cite{Jinping}.
	\item \textbf{\emph{$^9$Li/$^8$He}}: The spallation products $^9Li/^8He$ induced by cosmic-ray muons can mimic IBD signals through $\beta$-$n$ decays. The half-lifetimes of these isotopes are 173 ms/119 ms and the Q-value is 14 MeV/11 MeV. In general, $^9Li/^8He$ is generated near the correlative muons. Efficient muon track detection and time veto could suppress it.
	\item \textbf{\emph{Fast neutron}}:
	Energetic neutrons induced by cosmic ray muons can recoil protons or scatter with carbon nuclei which will give a prompt signal and followed neutron capture emits a gamma as a delayed signal. This can be largely suppressed by detector shielding. Furthermore, the prompt signal can be identified efficiently by particle identification with pulse shape discrimination (PSD) ~\cite{slow}.
	\item \textbf{\emph{Atmospheric neutrinos through a charged current (CC) interaction}}: High energy $\nu_{\mu}/\bar{\nu}_{\mu}$ can produce muons and delayed neutrons which mimic IBD signals. The atmospheric $\bar{\nu}_e$ background through the CC interaction is irreducible. However, the CC background induced by $\nu_{\mu}/\bar{\nu}_{\mu}$ is dominant.
	This background becomes more important with increasing energy.
	For boron neutrinos it is negligible.
	\item \textbf{\emph{Atmospheric neutrinos through a neutral current (NC) interaction}}: High energy atmospheric neutrinos may collide with $^{12}$C and eject a neutron which will imitate an IBD signal similarly to the fast neutron background.
	As demonstrated in~\cite{Randolph}, this background can be significantly decreased after searching for a coincident signal from the decay of a final isotope and using PSD.
	\item \textbf{\emph{Accidental coincidence and $^{13}$C($\alpha, n$)$^{16}$O}}: These two kinds of backgrounds depend on radioactivity. In addition, their spectra are in the low energy range, below 4 MeV and 6 MeV, respectively. For RE and LBE, they are negligible in comparison with other backgrounds~\cite{juno,acc_anbkg}.
\end{itemize}
Besides all above mentioned backgrounds, the thermal neutrino flux~\cite{Vitagliano:2017ona} from the Sun and the solar atmospheric neutrinos~\cite{Ng:2017aur} may contribute to electron antineutrinos but mostly in different energy orders, keV and GeV respectively, in comparison with typical IBD energies. 
The solar thermal neutrino flux comes from the electromagnetic interaction of the particles constituting plasma. 
The solar atmospheric neutrinos, in na\"{i}ve expectation, have as a similar production process as the atmospheric neutrinos on the Earth.~\cite{earth:atom_neu}.

Since an LBE is significantly distant from any nuclear power plants, we can use 3.4 MeV to 15.8 MeV as the energy window.
The lower bound of this interval excludes geo-neutrinos, which mostly contribute below 3.4 MeV~\cite{Jinping}.
We also consider another energy cut from 8.5 MeV to 15.8 MeV, which is above most of the reactor background.
The choice of lower bound (8.5 MeV) is based on~\cite{kamland_limit}.
In this window the background consists of atmospheric NC events~\cite{slow} and the tail of reactor spectrum.
It should be noted that the tail may have quite a large uncertainty. As can be seen from our calculations later in this paper, if the total background uncertainty is larger than 10\%, it is not possible to improve the current KamLAND result.
Our background is unbinned, i.e. there are no bins inside chosen energy windows.
Results for background estimation are based on the previous simulations from the JUNO and Jinping collaborations and summarized in Table~\ref{tab_1}.
Especially for the simulation of atmospheric NC background, GENIE was used for neutrino interactions~\cite{genie}.
\begin{table}[tbp]
\centering
\renewcommand{\arraystretch}{1.1}
\begin{tabular}{|c|c|c|c|c|c|}
\hline
Setup & LSc & kt$\cdot$year & 3.4-15.8 MeV (44.2 \%)& 8.5-15.8 MeV (13.8 \%)& 12-15.8 MeV (1.3 \%)\\
\hline
LBE & Slow & 10 & 136 & 0.25 & --\\
\hline
LBE & Normal & 10 & 148 & 13 & --\\
\hline
RE & Normal & 10 & -- & 65.5 & 0.2\\
\hline
\end{tabular}
\caption{\label{tab_1} The expected level of background in LBE  and RE with exposure time 10 $\rm kt\cdot year$. Two types of LSc are assumed~\cite{slow}. Fraction of boron spectrum for each window is shown in round brackets.}
\end{table}
\begin{figure}[ht]
\centering
\begin{minipage}[t]{0.47\linewidth}
\centering
\includegraphics[scale=0.38, right]{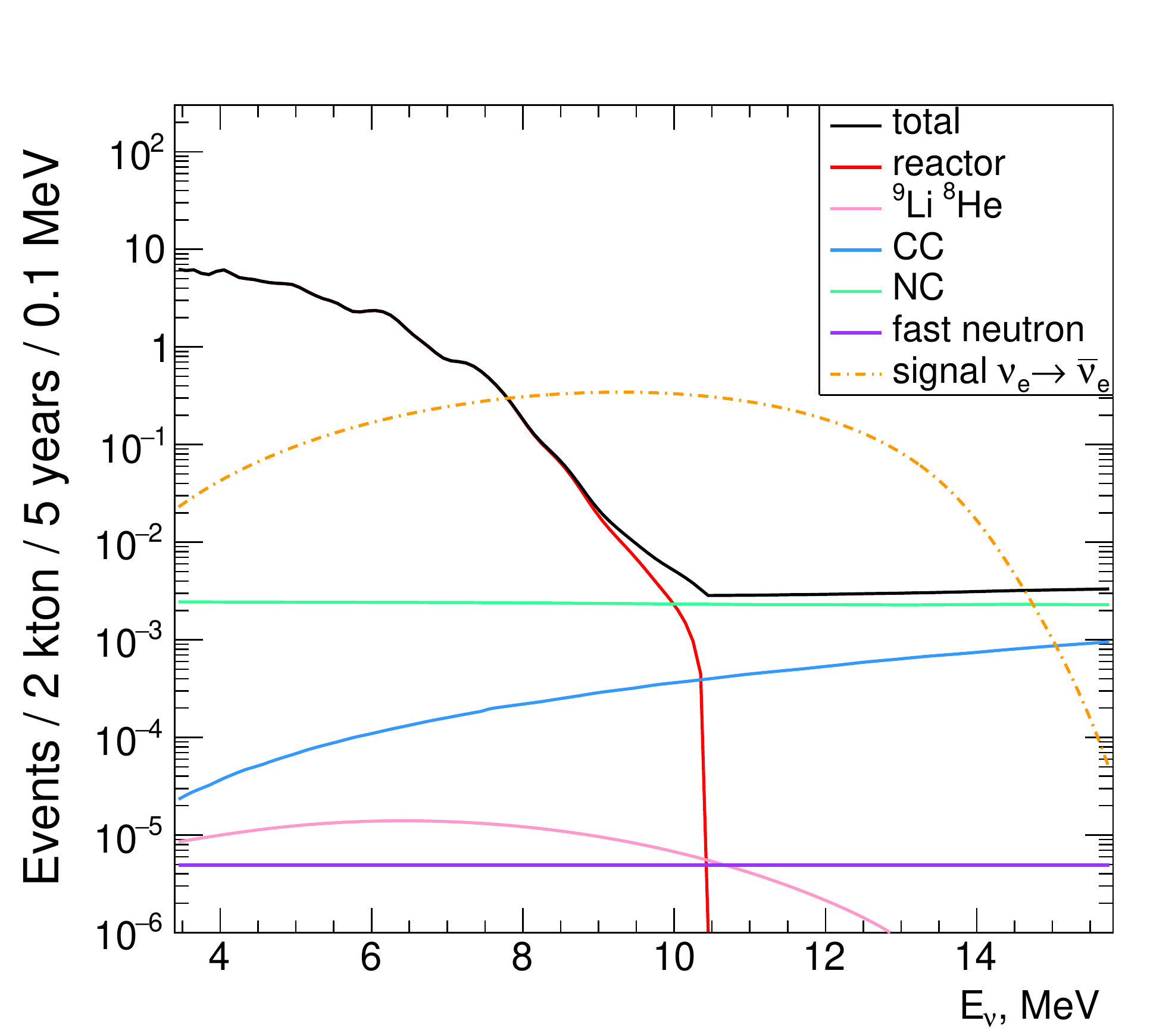}
\end{minipage}
\quad
\begin{minipage}[t]{0.47\linewidth}
\centering
\includegraphics[scale=0.38, right]{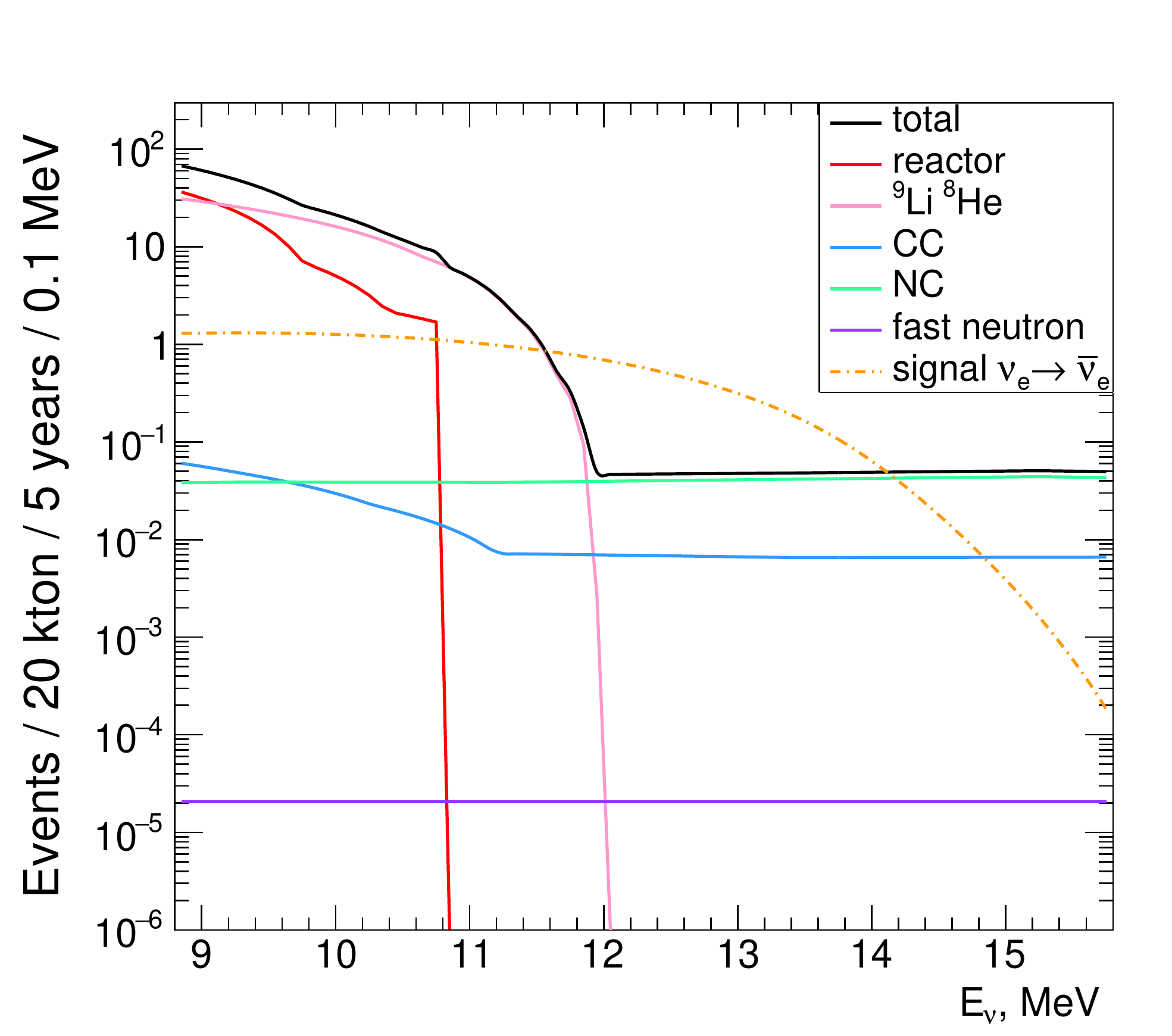}
\end{minipage}
\caption{\label{fig:2}The expected $\nu_e\rightarrow\bar\nu_e$ signal spectrum in the presence of the IBD-backgrounds.
The left panel corresponds to a LBE with slow LSc (exposure time 10 kt$\cdot$year), which can be associated with the Jinping detector and the right panel corresponds to a RE with normal LSc (exposure time 100 kt$\cdot$year), which can be associated with the JUNO detector.
The largest energy windows are chosen in both cases.
The signal is normalized to the upper limit of the averaged probability with the appropriate exposure time.}
\end{figure}

For RE the reactor background is much larger than for LBE, so it is reasonable to shift our energy windows.
This background with new energy windows is shown in the last row of Table~\ref{tab_1}.
It should be emphasized that all values for NC background are presented after appropriate treatment as described in~\cite{Randolph}.
\textcolor{red}{}
Approximation of the expected upper limits of signal and IBD-background for both types of experiments are shown in Figure~\ref{fig:2}.
Here the signal corresponds to the constant neutrino-antineutrino transition probability for the boron energy spectrum.
We assume an IBD selection efficiency ($\varepsilon$) of 80\% and 90\% for RE and LBE respectively with a relative error of 2\% in both cases.

\section{Analysis and results}
Our analysis examines the capabilities of measuring the average probability of neutrino-antineutrino transition~\eqref{eq_1} for both LSc experiment configurations.
For this we need to accurately estimate IBD-background within the chosen energy windows.
Let's assume, that the number of counts inside the detector is Poisson distributed and therefore its probability density function is:
\begin{equation}
\label{eq_2}
{\rm pdf}(n|\mu_{s})=\frac{(\mu_s+b)^n}{n!}\cdot \exp[-(\mu_s+b)],
\end{equation}
where $n$ -- number of observed events; $b$ -- expected background; $\mu_s$ -- average value of estimated signal.
Under this assumption background $b$ is known with some precision, hence it can be represented with a Gaussian constraint. An analogous constraint is used for $\mu_s$. Thus the equation~\eqref{eq_2} will take the following form:
\begin{equation}
\label{eq_3}
{\rm pdf}(n|\mu_{s})=\frac{(\mu_s^*+b^*)^n}{n!}\cdot \exp[-(\mu_s^*+b^*)]
\cdot\frac{\exp[-\alpha^2/(2\sigma_{\rm c}^2)]}{\sqrt{2\pi}\sigma_{\rm c}}\cdot\frac{\exp[-\beta^2/(2\sigma_{\rm \scriptscriptstyle BKG}^2)]}{\sqrt{2\pi}\sigma_{\rm \scriptscriptstyle BKG}},
\end{equation}
$$\mu_s^*=\mu_s(1+\alpha);\quad b^*=b(1+\beta),$$
where $\alpha$, $\beta$ -- nuisance parameters for estimated signal and background respectively; $\sigma_{\rm c}$ -- combined relative error of the signal; $\sigma_{\rm \scriptscriptstyle BKG}$ -- relative error of background.
For $\sigma_{\rm c}$ the dominant contribution comes from flux, so this value (12\%) is used in the most cases of the current analysis.
However for the energy window 12-15.8 MeV the shape uncertainty of the boron spectrum also contributes to the $\sigma_{\rm c}$.
This cannot be ignored, and for this uncertainty we assign $\sigma_{\rm c}=13.6\%$.
Usually in neutrino experiments background is well-known with relative error below 10\%~\cite{Randolph}, however in this analysis we decided to vary the $\sigma_{\rm \scriptscriptstyle BKG}$ up to 40\% and investigate what influence this has on the result.
Since we are interested in $\mu_s$ and this parameter is unknown, then it is reasonable to equate $n$ to the expected value of background $b$ and to evaluate the upper limit for $\mu_s$.

For statistical analysis we use {\sf RooStats}, which is based on {\sf RooFit}~\cite{roostats_1,roostats_2}.
This software provides excellent tools for interval estimations and hypotheses tests using different statistical methods without extensive coding.
Here we are using {\sf RooStats} to find the upper limit for parameter $\mu_s$.
Function~\eqref{eq_3} can be used as a pdf in {\sf RooStats} calculators.
In the current analysis we use two calculators, ProfileLikelihood and Bayesian.
The ProfileLikelihood calculator is based on the likelihood ratio, so-called profile likelihood function which is used to find a confidence interval~\cite{roostats_2}.
Standard definition of the profile likelihood function is:
\begin{equation}
\label{eq_4}
\lambda(\mu_s)=\frac{{\cal L}(n|\mu_s,\hat{\hat\alpha},\hat{\hat\beta})}{{\cal L}(n|\hat\mu_s,\hat\alpha,\hat\beta)},
\end{equation}
where the likelihood function in the numerator should be minimized with respect to nuisance parameters $\alpha,\beta$, and the parameter of interest (POI) $\mu_s$ is fixed.
The likelihood function in the denominator should be minimized with respect to all parameters.
The one-side 90\% C.L. for the POI can be obtained from $-\ln(\lambda)=1.645$.
The Bayesian calculator computes the posterior probability of the POI.
We use a Gaussian distribution as the prior distribution for parameters $\alpha$ and $\beta$.
For  this calculator the one-side 90\% C.L. for the POI is determined as 90\% of the area under posterior distribution.
Both calculators are in good agreement for large statistics (more than 10 counts).
However for low statistics we use the Bayesian calculator, because in this case the upper limit is independent of background.

\begin{figure*}[h!]
\centering
\begin{minipage}[t]{0.47\linewidth}
\centering
\includegraphics[scale=0.3]{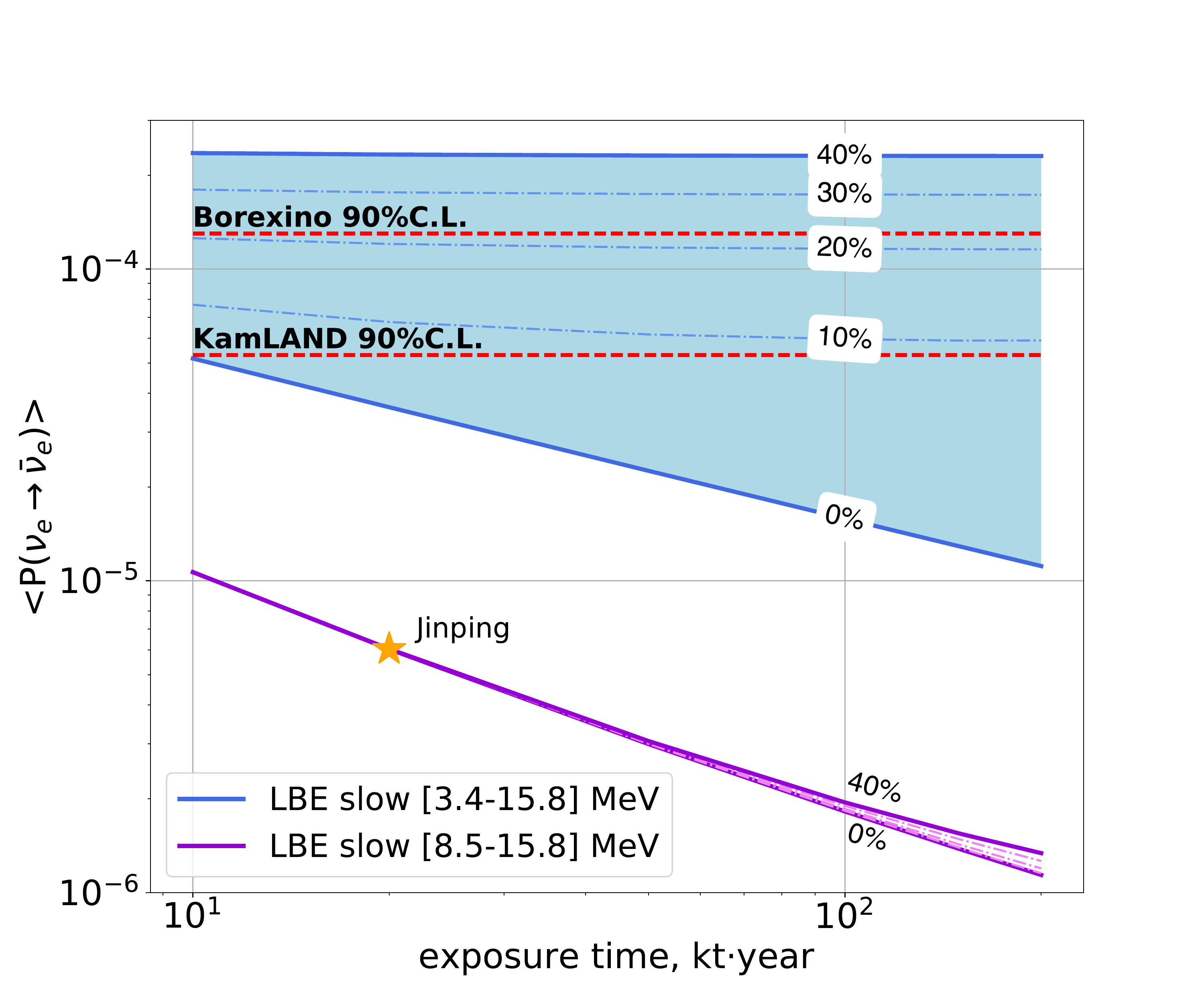}
\end{minipage}
\qquad
\begin{minipage}[t]{0.47\linewidth}
\centering
\includegraphics[scale=0.3]{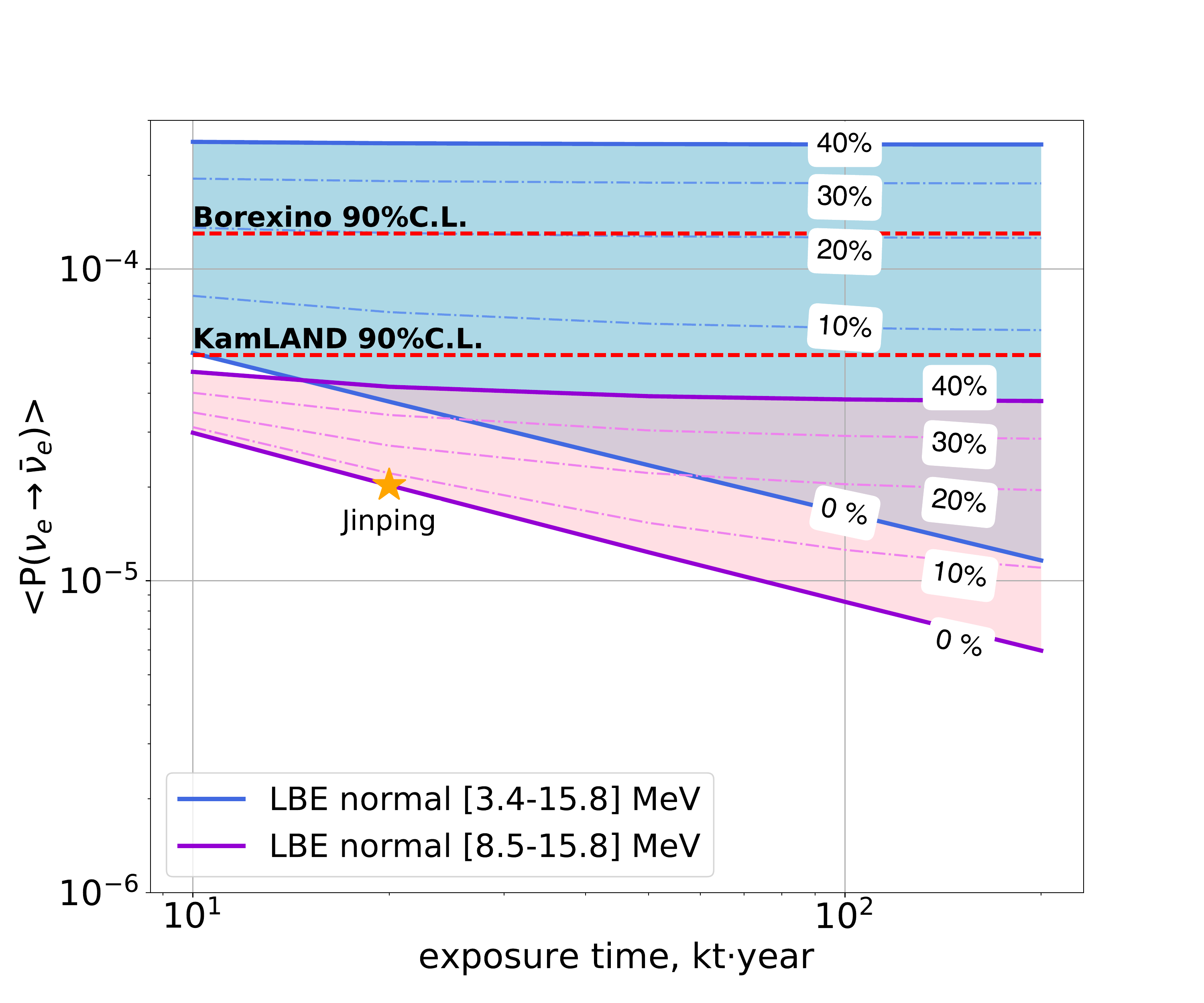}
\end{minipage}
\vfill
\begin{minipage}[b]{0.5\linewidth}
\centering
\includegraphics[scale=0.3]{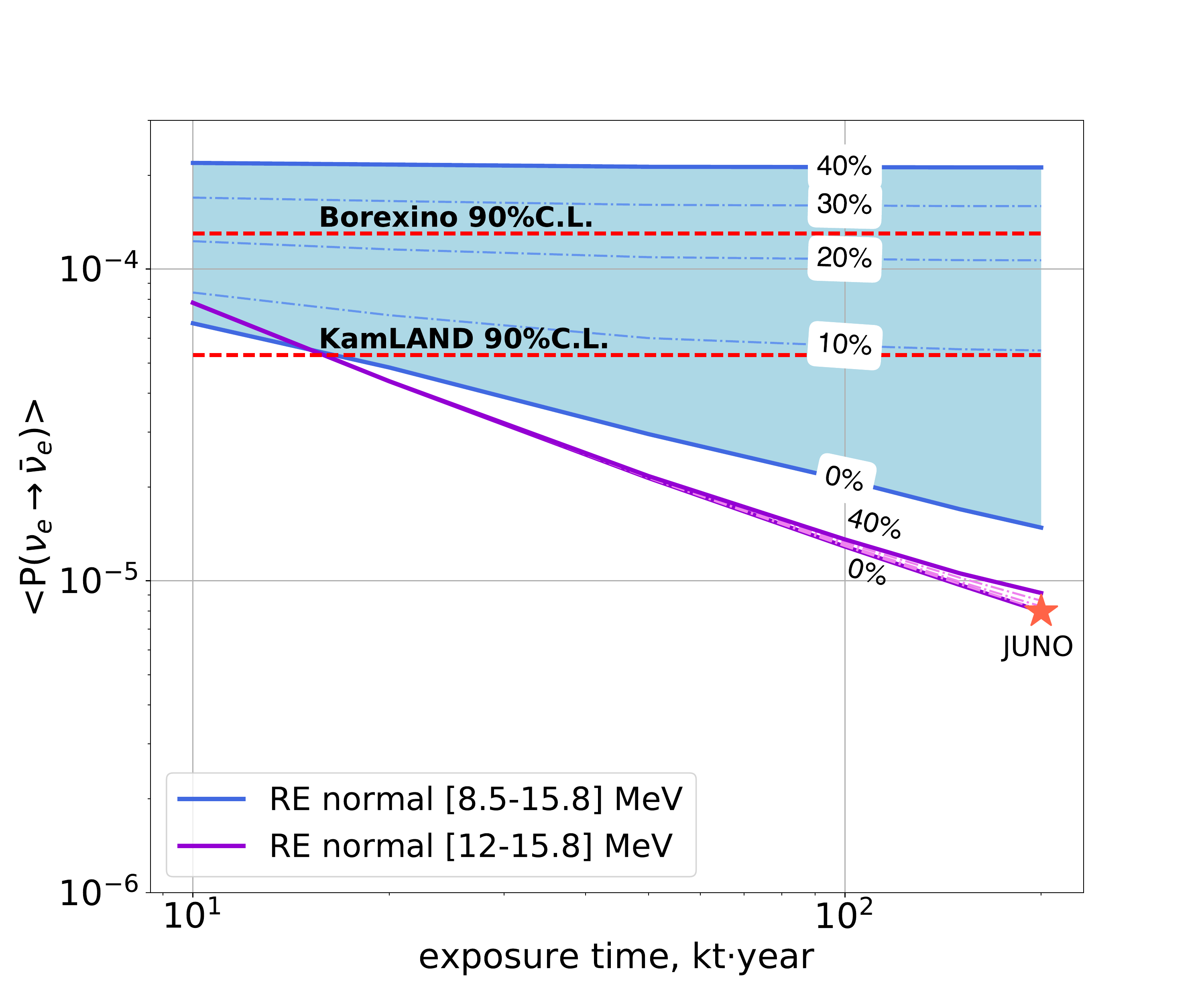}
\end{minipage}
\caption{\label{fig:4}The 90\% C.L. for average transition probability between neutrinos and antineutrinos in different energy windows for different values of background uncertainty from 0\% to 40\% as a function of exposure time. The top left panel corresponds to an LBE with slow LSc and the top right panel -- LBE with normal LSc. The bottom panel corresponds to the RE case with normal LSc. The red dashed lines show the current experimental limit from Borexino and KamLAND.}
\end{figure*}

The 90\% C.L. for average probability~\eqref{eq_1} as a function of exposure time is shown in Figure~\ref{fig:4}.
Additionally, our calculations demonstrate the advantage of using an energy window with lower background as opposed to higher statistics over a larger range of energy.
Furthermore for large energy windows increasing exposure time does not improve the results, if the background error is higher than a few percent.
It should be emphasized that the optimization of energy window should not include the compression of  their width, because it will cause the denominator in formula~\eqref{eq_1} to decrease and consequently increase the probability.
Both experimental facilities can improve the current experimental limit for neutrino-antineutrino transitions by at least an order of magnitude, however it is necessary to choose the energy windows carefully with focus on lower background across the window.

\section{Conclusion}

The new generation of LSc detectors will allow us to improve our understanding of the 3-neutrino oscillation paradigm, but also look for new physics.
Transition between neutrinos and antineutrinos is a promising area in which to look.
The observation of this phenomenon will at minimum confirm the breaking of CPT symmetry and violation of total lepton number by two units.
Boron neutrinos could be a convenient candidate to search for such transitions due to their well-predicted spectrum and the fact that they have the longest propagation time among of all terrestrial neutrino (antineutrino) sources.

This research has demonstrated the superiority of using small energy windows with low background, especially when background is very close to zero, in comparison with large windows.
An LBE with large mass and exposure time is better choice for this measurement in comparison with RE.
Using real data it will be possible to optimize the width of the energy windows and thus slightly improve sensitivity.
If transitions between neutrinos and antineutrinos exist with probability greater than $6\times10^{-6}$ and $9\times10^{-6}$ for reference points Jinping and JUNO respectively, then both experimental facilities have a big chance to observe this phenomenon during an exposure time of 10 years.
Additionally the results of this count calculations can be applied to other exotic searches like antineutrinos from dark matter annihilation or antineutrinos from astronomic sources.

\section*{Acknowledgments}
This work was in part supported by the China Postdoctoral Science Foundation under Grant No. 2018M643283,
the National Key R\&D program of China under Grant No. 2018YFA0404103 and Key Lab of Particle \& Radiation Imaging, Ministry of Education.
We would like to say great thanks to Prof. Wang Zhe and Dr. Li Yufeng for providing information about background in Jinping and JUNO respectively.
We express our special gratitude to Prof. Cao Jun for his important, valuable remarks and corrections.


\end{document}